# Disentangling the dominant drivers of gravity wave variability in the Martian thermosphere

N. V. RAO[1*], V. LEELAVATHI[2], CH. YASWANTH[1], S. V. B. RAO[2]

[1]National Atmospheric Research Laboratory, Gadanki, India
[2]Department of Physics, S. V. University, Tirupati, India

[*]*Correspondence to: nvrao@narl.gov.in*

**ABSTRACT**

In this study, we extracted the amplitudes of the gravity waves (GWs) from the neutral densities measured in situ by the neutral gas and ion mass spectrometer aboard the Mars atmosphere and volatile evolution mission. The spatial and temporal variabilities of the GWs show that solar activity (the F10.7 cm solar flux corrected for a heliocentric distance of 1.66 AU), solar insolation, and the lower atmospheric dust are the dominant drivers of the GW variability in the thermosphere. We developed a methodology in which a linear regression analysis has been used to disentangle the complex variabilities of the GWs. The three dominant drivers could account for most of the variability in the GW amplitudes. Variability caused by the sources of GWs and the effects of winds and the global circulation in the mesosphere and lower thermosphere are the other factors that could not be addressed. The results of the present study show that for every 100 sfu increase in the solar activity, the GW amplitudes in the thermosphere decrease by ~9%. Solar insolation drives the diurnal, seasonal and latitudinal variations of ~9%, ~4% and ~6%, respectively. Using the historical data of the dust opacity and solar activity, we estimated the GW amplitudes of the Martian thermosphere from MY 24 to MY 35. The GW amplitudes were significantly reduced during the maximum of solar cycle 23 and were highest in the solar minimum. The global dust storms of MY 25, 28, and 34 lead to significant enhancements in the GW amplitudes.

## 1. INTRODUCTION

Gravity waves (GWs) are small-scale disturbances in atmospheric variables such as pressure, density, wind, and temperature. Though the waves are ubiquitous in atmospheres of the terrestrial planets (Earth, Mars and Venus), their amplitudes are expected to be stronger on Mars due to its relatively tenuous atmosphere. For example, recent air pressure measurements by Insight and Curiosity have shown that Mars surface atmosphere is rich in gravity waves (Banfield et al. 2020; Guzewich et al. 2021), which are likely generated by wind flow over topography. In fact, topography is the main source of GWs with small observed phase velocities on Mars (Pickersgill & Hunt 1979, 1981), though other sources such as convection can also produce considerable GW activity (Creasey et al. 2006; Spiga et al. 2013; Imamura et al. 2016). Recent studies have shown that the GW activity in the Mars lower atmosphere also varies with the lower atmospheric dust (Heavens et al. 2020; Kuroda et al. 2020; Guzewich et al. 2021). Using Curiosity observations, Guzewich et al. (2021) have reported that in the entire period of their observations (nearly four Martian years), the GW activity over Gale crater was strongest during the 2018 global dust storm (GDS). However, it is not known whether the variations observed at the Gale crater are region specific or present across the globe. Other studies have shown that the GW activity in the lower atmosphere overall decreases during the regional and global dust storms (Heavens et al. 2020; Kuroda et al. 2020). Thus, the exact role of the lower atmospheric dust in the generation or variability of GWs is not known. Furthermore, GWs over the Gale crater show diurnal variability with strongest wave activity in the evening and early morning hours (Guzewich et al. 2021). The wave activity also displays a notable seasonal variation with strongest activity in the southern spring and summer seasons (solar longitude, $L_s$ = 180 – 330°) with a local minimum in activity near the southern summer solstice ($L_s$ = 270°) (Guzewich et al. 2021). GWs observed at 20-30 km altitudes also display spatial and temporal variations that are likely related to variations at the sources of their generation (Heavens et al. 2020).

The characteristics of the GWs in the Mars upper atmosphere have been studied for several decades using measurements made by the Mars Global Surveyor, Mars Odyssey, Mars Reconnaissance Orbiter, the Mars Atmosphere and Volatile EvolutioN (MAVEN), and the Trace Gas Orbiter (Creasey et al. 2006; Fritts et al. 2006; Withers 2006; Tolson et al. 2007; Withers & Pratt, 2013; Yiğit et al. 2015; England et al. 2017; Terada et al. 2017; Vals et al. 2019; Jesch et al. 2019; Leelavathi et al. 2020). The GWs in the thermosphere display two important features that are distinctly different compared to those of the lower atmospheric GWs. First, amplitudes of the GWs in the thermosphere are anywhere between 5% and 50% of the background densities and temperatures (Yiğit et al. 2015; England et al. 2017; Terada et al. 2017; Vals et al. 2019; Rao et al., 2021), whereas those of the lower atmospheric GWs are mostly <1% (Guzewich et al. 2021). This is because the majority of the GWs observed in the thermosphere originate in the lower atmosphere and as these GWs propagate upward into the tenuous upper atmosphere, their amplitudes grow significantly resulting in larger amplitudes in the thermosphere. However, the amplitudes of the waves cannot grow indefinitely and the state of the underlying atmosphere regulates their growth. This leads to the second important feature that the variability of thermospheric GWs is dependent more on the state of the underlying atmosphere rather than the variability at their sources. It has been shown that the amplitudes of the GWs in the thermosphere are inversely related to the local temperatures (Yiğit et al. 2015; England et al. 2017; Terada et al. 2017; Rao et al. 2021). This inverse relation was explained by invoking the saturation of GWs due to convective instability in the thermosphere (Terada et al. 2017; Vals et al. 2019). However, Yiğit et al. (2021b), using a one-dimensional spectral nonlinear GW model, have argued that the role of the convective instability mechanism is likely smaller in producing the day-night difference in GW activity. Starichenko et al. (2021) have also come with nearly similar conclusions for the growth of the GWs in the middle atmosphere.

When the temperature dependency was removed, Terada et al. (2017) found that the GW amplitudes do not show any significant variation with geographic location and solar wind parameters. This suggests that the major variability in GW amplitudes is related to the background temperatures. Though the exact role of the lower atmospheric dust in the variability of GWs in the lower atmosphere is not yet firmly established, recent findings have shown that in the thermosphere the amplitudes of the GWs are enhanced during GDSs (Leelavathi et al. 2020). Such an enhancement was explained by surmising that the GDS modified the winds and circulation in the mesosphere and lower thermosphere in such a way that the background conditions are more conducive for the upward propagation of the GWs. However, the possibility of GDS generating GWs in the lower atmosphere cannot be ruled out, particularly during the period of the deep convective activity (e.g., Imamura et al. 2016).

Thus, the previous studies suggest that thermospheric temperatures and the lower atmospheric dust are the two major controlling factors of the thermospheric GWs. Quantifying the contributions of these factors to the thermospheric GW variability is useful to better understand the Mars upper atmosphere and its gaseous escape. For example, 20-40% increase in GW amplitudes in the Mars thermosphere is shown to enhance the net escape flux of hydrogen by ~1.3-2 times through thermal escape mechanism (Yiğit et al. 2021a). However, disentangling the complex variabilities of the GW amplitudes and quantifying the contributions of the thermospheric temperatures and the lower atmospheric dust to the variability of the GW amplitudes is a difficult task due to the inherent complexity of the spacecraft measurements. The spacecraft measurements, from which GW amplitudes are extracted, are generally made at varying latitudes and local solar times (LSTs) resulting in coarser spatial and temporal resolutions of the data. Therefore, several months to years of data are required to properly understand these variabilities. Inter-annual variabilities and other long-term variabilities further complicate the problem. In this work, we developed a methodology that successfully isolates the contributions of the thermospheric temperatures and lower atmospheric dust to the GW amplitudes in the Martian thermosphere. Considering the fact that solar heating is the main driver of the thermospheric temperatures, the analysis is further extended to quantify the contribution of the solar activity to the GW variability.

## 2. INSTRUMENT, DATA AND METHODOLOGY

### 2.1. NGIMS Data and Coverage

MAVEN was placed in the Martian orbit in September 2014 in a highly eccentric orbit with periapsis and apoapsis altitudes of ~150 km and ~6,220 km, respectively. The orbital period and inclination of MAVEN are ~4.5 h and 75°, respectively (Jakosky et al. 2015). Occasionally, MAVEN's periapsis has been brought down to altitudes as low as ~120 km (Bougher et al. 2015; Jakosky et al. 2015) that allow the observation of the Martian lower thermosphere. Such "deep dip (DD)" campaigns usually last for a week and there were nine such campaigns during the period of the observations used in this study. Data from these observations is not included in the present study (Leelavathi et al. 2020). MAVEN carries a suite of instruments that measure the particles and fields in the Martian upper thermosphere, exosphere and in its near space environment. One such instrument is the Neutral Gas and Ion Mass spectrometer (NGIMS), which measures both the neutral and ion species in the Mars upper atmosphere.

The present study uses the argon (Ar) densities measured in situ by the NGIMS instrument onboard the MAVEN spacecraft. NGIMS is a dual-source quadrupole mass spectrometer designed to measure the Martian upper atmospheric neutral and ion densities in the m/z range of 2–150 amu, with unit mass resolution (Mahaffy et al. 2015a). The instrument is operated in the closed as well as in open source modes. In the closed source mode, NGIMS measures the nonreactive neutral species and this mode of operation is carried out in all orbits. In the open source mode, the reactive neutrals and ions are measured in alternate orbits. The present study uses Ar densities measured in the closed source mode. In general, NGIMS is operated when MAVEN is in its periapsis pass (below 500 km in the inbound segment, through the periapsis and until the spacecraft moves above 500 km in the outbound segment) (Benna et al. 2015; Mahaffy et al., 2015a, 2015b). In the present study, the amplitudes of the GWs are extracted from the Ar density profiles (given in section 2.2). Such an extraction needs sufficient signal of Ar to determine the perturbations effectively above the noise level of the instrument. Moreover, Ar density profiles often do not extend above 220 km. Considering these limitations, the GW extraction in the present study is restricted for altitudes < 200 km. In addition, we use data from only the inbound segment of each orbit.

The present study uses the Level 2, version 08, revision 01 data of NGIMS database. The data spans from 21 February 2015 (solar longitude, Ls=295° in Martian year (MY) 32) to 15 June 2020 (Ls=220° in MY 35). The thermospheric temperatures are derived (section 2.3) from the Ar density profiles. In addition, F10.7 cm solar flux is used as a proxy for the solar activity. Regularly kriged 9.3 µm absorption column dust optical depth ($\tau$) at 610 Pa, taken from the Mars Climate Database (Montabone et al. 2020), are used as a proxy for the lower atmospheric dust activity.

**2.2. Extraction of the Gravity Wave Amplitudes**

The method of extraction of the amplitudes of GWs is similar to that used in a previous study by Leelavathi et al. (2020). At first, an inbound portion of a density profile is fitted with a seventh order polynomial fit to estimate the background density (ρ). Density perturbations (δρ) are estimated by removing the background density profile from the measured density profile. Relative density perturbations (δρ/ρ) are obtained by normalizing the density perturbations with the background densities. These perturbed densities are further subjected to the Lomb-Scargle periodogram (Lomb 1976; Scargle 1982) to identify the presence of GWs. From each orbit, we consider only the dominant GW and estimate its amplitude and wavelength. From the estimated amplitude and wavelength, we reconstructed the perturbations. A comparison between the original and reconstructed perturbations confirms that the method of extraction of GWs adopted in this study captures the maximum perturbation in each density profile (Leelavathi et al. 2020). Note that close to the periapsis the spacecraft covers both the vertical and horizontal distances. Within the altitude range considered in this study, however, the horizontal displacement of the spacecraft is several times that of the vertical. We, therefore, consider the extracted waves as the horizontal GWs.

**2.3. Extraction of the Thermospheric Temperatures**

The thermospheric temperatures are computed from the Ar density profiles using a method similar to that used by Snowden et al. (2013) for deriving the temperatures of the Titan's upper atmosphere and subsequently adopted for Martian studies by Stone et al. (2018) and Leelavathi et al. (2020). For each orbit, the temperature at the upper boundary is computed by fitting to the top of the atmosphere (between densities of $1\times10^4$ to $4\times10^5$ cm$^{-3}$), an equation of the form

$$N(r) = N_0 \cdot exp\left[\frac{GMm}{KT}\left(\frac{1}{r} - \frac{1}{r_o}\right)\right] \quad \text{------- (1)}$$

where $N_0$ is the density at the lower boundary of the fitted region and $r_o$ is the distance from the centre of the planet to the lower boundary of the fitted region, $r$ is the distance from the centre of the planet, $N$ the number density, $m$ is the mass of Ar, $G$ is the gravitational constant, $K$ is Boltzmann constant and *M* is the mass of Mars. The temperature extracted in this way is considered as the temperature of the upper boundary of the profile. The partial pressure at the upper boundary ($P_0$) is then computed using the ideal gas law, $P_0$= *NKT*. From the hydrostatic equation, the pressure at a given altitude is given by

$$P(r)=P_0+GMm \int_r^{r_{uo}} N(r)\frac{dr}{r^2} \quad \text{------ (2)}$$

Here, $r_{uo}$ is the distance from the centre of the planet to the upper boundary (corresponding to $P_0$). Considering the resultant pressure profile, the temperatures at each altitude are then obtained by using the ideal gas law. The thermospheric temperatures used in this study are obtained by averaging each temperature profile from the periapsis to 200 km.

## 3. RESULTS

Before studying the variability of thermospheric GWs, we first present the precession of the MAVEN spacecraft in LST and latitude (Figure 1a) and conditions of the solar flux and the lower atmospheric dust (Figure 1b) for the period of the observations used in this study. The intermittent gaps in the MAVEN observations (Figure 1a) are mainly due to the operation of the spacecraft for DD campaigns and due to the non-availability of observations during the solar conjunction periods and during the communication of the spacecraft with the Earth (Jakosky et al. 2015; Stone et al. 2018). As can be seen from the figure, the data cover a wide range of latitudes, seasons, and LSTs. From Figure 1a it can be noted that at least a few months of observations are required to get a full diurnal or latitudinal coverage. This is due to the slow precession of the MAVEN periapsis in LST and latitude.

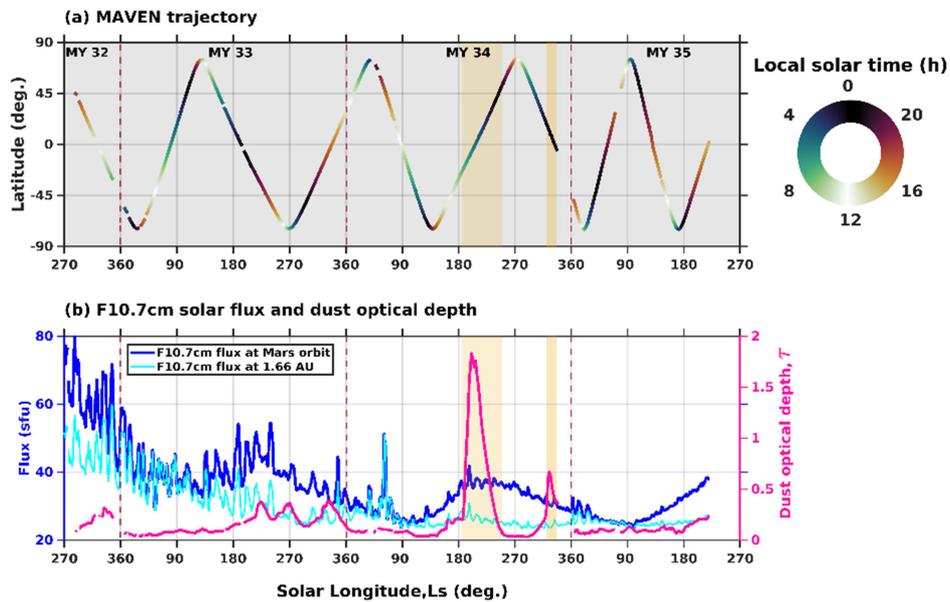

*Figure 1.* Temporal variation of (a) latitude of MAVEN trajectory color coded by the local solar time, and (b) F10.7 cm solar flux and the dust optical depth, τ. Solid blue curve shows the solar flux corrected for the Martian orbit and cyan curve shows the solar flux corrected for a constant heliospheric distance of 1.66 AU. Vertical red lines enclose different Martian years. Shaded regions are the periods of the global and regional dust storms in MY 34. Solar fluxes are shown in solar flux units (sfu). The τ values shown are obtained by averaging the τ values over all longitudes and between latitudes of ±10° around the MAVEN periapsis latitude footprint.

The solar F10.7 cm flux corrected for the Mars orbit (blue line) shown in Figure 1b indicates that the MAVEN observations used here were obtained in the descending phase of the solar cycle (from medium flux in MY 32 to minimum flux in MY 35). The solar flux also shows an annual variation due to the change in the Sun-Mars distance. The cyan line in Figure 1b presents the solar flux for a constant distance of 1.66 AU depicting the gradual decrease of the flux, from the beginning of observations to the solar minimum. In other words, this flux (cyan line) is a de-seasonalized term and is free from the variability due to Sun-Mars distance. This de-seasonalized F10.7 cm solar flux is hereafter referred to as 'solar activity' and is used throughout the rest of the study. From the dust optical depth, $\tau$ (magenta solid line in Figure 1b), we can note that the dust content in the Mars lower atmosphere increases in the second half of each MY. Notably, MY 34 marks the 2018 GDS with significant enhancement of $\tau$ between Ls = 185° and Ls = 240°. There was also a regional dust storm in MY 34 between Ls = 320° and Ls = 336°.

Figures 2a & 2b show the thermospheric GW amplitudes and temperatures, respectively as a function of Ls (averaged from all orbits that are within one degree of Ls). Under nominal dust conditions, the maximum amplitudes of GWs are ~15% and during the 2018 GDS the amplitudes reached as high as 22% of the background densities. The thermospheric temperatures lie between 100 K and 280 K. The GW amplitudes and temperatures show a sinusoidal variation that is primarily due to the LST variability (c.f., Figure 1a) (Leelavathi et al. 2020; Rao et al. 2021; Yiğit et al. 2021b). The nighttime GW amplitudes are not only larger but also have more spread (the range of amplitudes at a given Ls). Particularly during the 2018 GDS (Ls:185°-240°), the amplitudes of the GWs extend from ~8% to 22% showing the highest spread. The diurnal variation of the thermospheric temperatures (Figure 2b) during the nominal dust period is nearly opposite to that of the GWs. During the period of the GDS, however, an enhancement is observed both in the GW amplitudes and the thermospheric temperatures (Leelavathi et al. 2020).

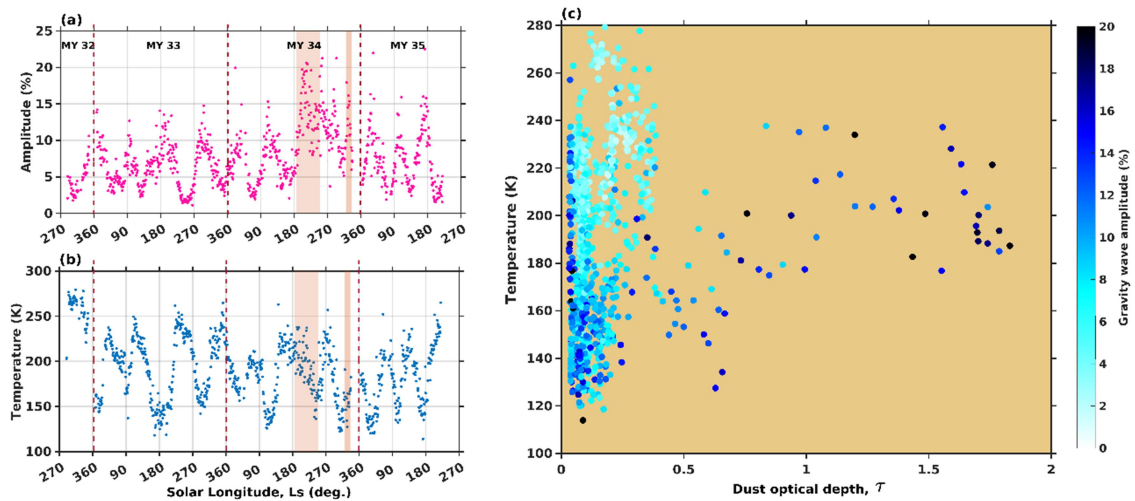

*Figure 2.* Temporal variation of the (a) amplitudes of GWs and (b) thermospheric temperatures. (c) Temperature and dust optical depth dependency of the amplitudes of GWs.

The relation of GW amplitudes to the thermospheric temperatures and to the dust optical depths is further studied in Figure 2c. Under low dust ($\tau < 0.4$) conditions, the GW amplitudes show opposite relation to the thermospheric temperatures, which is consistent with the previous reports (England et al. 2017; Terada et al. 2017; Vals et al. 2019; Siddle et al. 2019; Leelavathi et al. 2020; Rao et al. 2021). Under high dust conditions ($\tau > 1.4$), however, the amplitudes of GWs are clearly larger (Leelavathi et al. 2020). From this it appears that the thermospheric temperatures and the lower atmospheric dust solely can account for most of the variability in GW amplitudes. In the following, a simple linear regression analysis using an equation of the form

$$Q(T_{th}, \tau) = 18.48(\pm 1.07) - 0.06 \pm (0.01)T_{th} + 6.09(\pm 0.74)\tau \text{ ---(3)}$$

is used to quantify the response of the GW amplitudes to the thermospheric temperatures and to the lower atmospheric dust content. Here, $Q$ represents the amplitudes of the GWs observed in the thermosphere, $T_{th}$ is the thermospheric temperature. The values in the brackets are uncertainties in the estimation of the coefficients. The coefficients of the regression suggest that for a 100 K increase in the thermospheric temperatures, the GW amplitudes are reduced by ~6% and for one unit increase in $\tau$, the GW amplitudes are increased by ~6%.

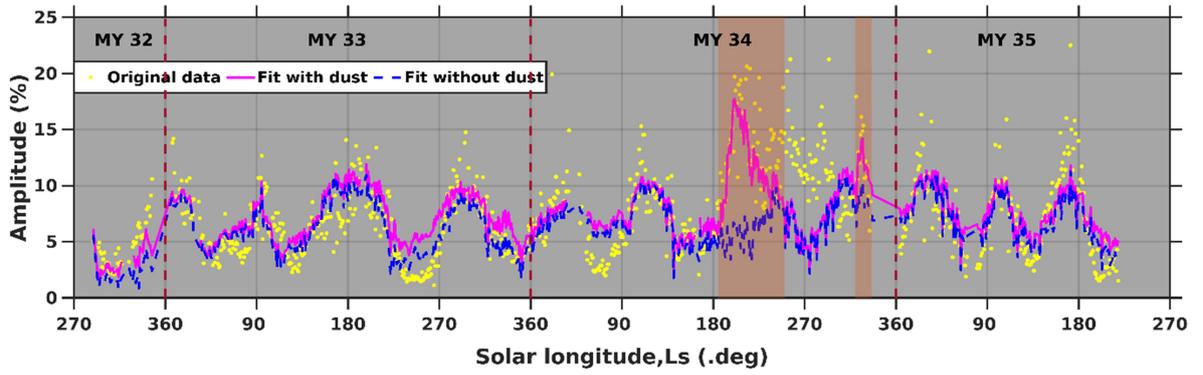

*Figure 3: Temporal variation of (yellow filled circles) GW amplitudes along with the best fit curves (solid) with the dust contributions included and (dashed) without the dust contribution. The shaded regions enclose approximate periods of the global (Ls=185°-240°) and regional dust storms (Ls=320°-336°) in MY 34.*

Figure 3 shows the original GW amplitudes and the curves fitted using equation 3. Here, two fits are shown: one with dust and the other without the dust component. The root mean square (RMS) values of the original data, the fit without dust, and the fit with dust are 8.45, 6.72, and 7.92, respectively. This shows that the fit without dust can account for ~79.5% of the variability in the original data. With inclusion of the dust, the fit accounts for ~93.7% of the variability in the original data. Except during the periods of the 2018 GDS and the regional dust storm, the two fitted curves agree reasonably well and go hand-in-hand. During the dust storm periods, however, the fit that includes the dust shows better agreement with the observations. For example, at Ls=210° in MY34, the fit without dust predicts an amplitude of ~ 6% whereas the fit that includes dust predicts the amplitudes to be ~18%. Thus, the GDS increases the GW amplitudes by ~ 12% at the peak of the storm. During the GDS period, the RMS values of the original data, the fit without dust and the fit with dust are 14, 6.58, and 12, respectively. Thus, during GDS the fit without dust accounts for only 47% of the variability whereas the fit with dust accounts for 85.7% of the variability. Note that in MY 33 also, the fit with dust predicts somewhat more GW amplitudes during Ls=220°-275° and Ls=320°-340°, compared to the one without dust. These two periods in MY 33 are also associated with enhanced dust content (c.f., Figure 1b). Note that both the fitted curves deviate from the original values when the GW amplitudes are at their minimum or maximum in each diurnal cycle. A large deviation of both the fitted curves is also observed after the peak of the GDS in MY 34 (Ls=210°-300°). This aspect is discussed in detail in the following while dealing with Figure 5.

In the following, we try to quantify the contribution of the solar activity to the GW amplitudes. At first place, the GW amplitudes and the solar activity appear unrelated. However, the fact that the solar EUV and X-ray heating plays a major role in driving the temperatures of the Mars upper atmosphere (e.g., Forbes et al. 2006; Bougher et al. 2015) connects the GWs and solar activity in an indirect way. In fact, MAVEN observations have shown an increase in GW amplitudes with decrease in solar activity from medium to low levels of solar irradiance (Rao et al. 2021). Yiğit et al. (2021b) have also reported larger amplitudes of the GWs during the recent solar minimum. Quantifying the GW variability in terms of solar activity has the added advantage of predicting the GW variability with the past data of the solar activity. Therefore, in the following we quantify the contribution of the solar activity in driving the GW amplitudes. However, it is known that the planet's eccentricity, obliquity and rotation lead to uneven distribution of the solar insolation on the Mars atmosphere and lead to the seasonal, latitudinal and diurnal variations, respectively. Therefore, before attempting to isolate the solar activity contribution, we first examine the observed GW amplitudes for any signatures of the solar insolation effects.

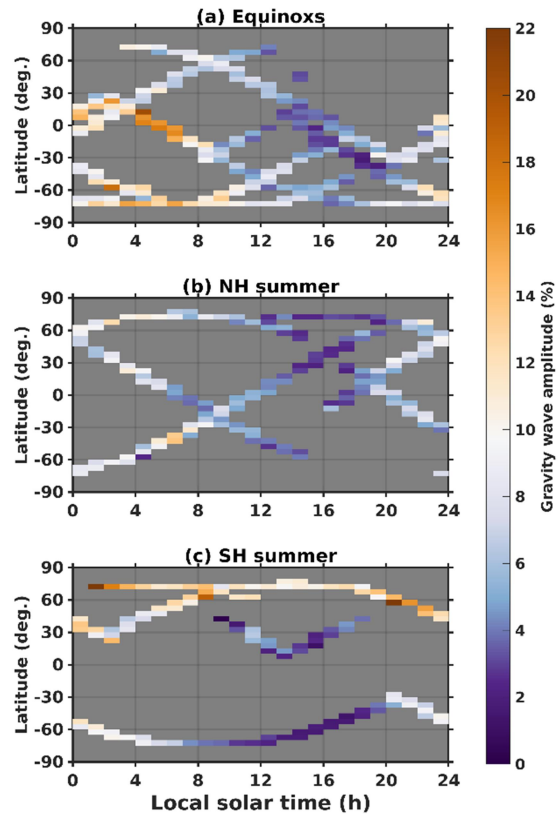

*Figure 4*: LST and latitude variation of the amplitudes of gravity waves for (a) equinoxes (Ls=0° and Ls=180° combined), (b) Northern Hemispheric (NH) summer (Ls=90°), and (c) Southern Hemispheric (SH) summer (270°). Each season covers an Ls range of ±45° centering its respective seasonal cardinal point. The latitudes and LSTs correspond to the MAVEN periapsis footprints. The GW amplitudes are averaged over 1 h in LST and 5° in latitude. LST= Local solar time.

Figure 4 shows the LST vs. latitude variability of the GW amplitudes in different seasons; equinoxes (Ls=0°±45° & Ls=180°±45° combined), northern hemispheric (NH) summer (Ls=90°±45°), and southern hemispheric (SH) summer (Ls=270°±45°). Here data from the two equinoxes are combined since the effects of the solar insolation in equinoxes are expected to be symmetric around the equator. In addition, combining data from the two equinoxes increases the spatial coverage so that the latitudinal and LST features can be clearly discerned. The diurnal variability of the GWs with smaller amplitudes during afternoon hours and larger amplitudes in the post-midnight is consistently observed in all the seasons. In equinoxes, the GW amplitudes are larger at high-latitudes than over the equator, except between 4-7 h where an enhancement is observed over the equator. This early morning enhancement in GW amplitudes over the equator in equinoxes was due to the 2018 GDS (Leelavathi et al. 2020). No significant latitudinal variability is observed in NH summer, which is close to the aphelion (Ls=71°). In SH summer, however, the amplitudes are larger in the winter hemisphere than in the summer hemisphere. Note that the SH summer is close to the perihelion (Ls=251°) where the solar insolation is strongest. Thus, from Figure 4, it appears that the GW amplitudes show solar insolation variability with larger amplitudes in regions of low solar insolation and vice versa. Note that SH summer is also the season with high dust content in the lower atmosphere. Hence, it should be noted that the GWs shown in Figure 4c contain contributions from the dust as well. In general, the eccentricity of the ecliptic, obliquity of a planet and its rotation cause the seasonal, latitudinal and diurnal variabilities in the solar insolation. The variabilities in GW amplitudes caused by the solar activity, the lower atmospheric dust and the solar insolation can be isolated through regression analysis using an equation of the form

$$Q(F_{10.7}, D, \lambda, \delta, t) = Q_0 + Q_1 F_{10.7} + Q_2 \tau + Q_3 \left\{ \frac{[\sin(\lambda) \sin(\delta) + \cos(\lambda) \cos(\delta) \cos 15(t-14.5)]}{\left(\frac{\max(r)}{r}\right)^2} \right\} \quad \text{--- (4)}$$

Where the declination, $\delta = \varepsilon \sin(L_s)$

and $r = \frac{r_m(1+e)}{1+(e \cos(L_s + 15(t-14.5) - 81))}$

here $F_{10.7}$ is the solar activity, $\lambda$ is the latitude of observations, $\varepsilon$ is the Mars obliquity, 't' is the LST, and 'e' is eccentricity of the planet. 'r' is a variable that incorporates both the Sun-Mars distance and LST and $\max(r)$ is the maximum value of 'r'. The values of coefficients $Q_0$, $Q_1$, $Q_2$, and $Q_3$ shown in the above equation are 9.61±0.84, −0.09±0.03, 3.57±0.67, and −5.23±0.38. These are the best fit coefficients obtained through a linear regression analysis. The value following each best fit coefficient is the uncertainty in the estimation of that coefficient. Note that to obtain the solar activity coefficient, we used the F10.7 cm solar flux corrected for a constant heliospheric distance of 1.66 AU (shown in Figure 1b). The variations related to Sun-Mars distance are accounted for by the solar insolation (the fourth) term of equation (4). Note that we often refer to the solar activity and solar insolation contributions while dealing with the nightside observations. This is due to the fact that the daytime temperatures are transported to the nightside, where their effects can be felt by the upward propagating GWs.

The reconstructed GW amplitudes from equation (4) are shown in Figure 5a. For comparison, the original values are also shown. Contribution of the dust to the thermospheric GW amplitudes can be better appreciated by comparing the solid (reconstructed with the dust contribution) and dashed (reconstructed without the dust contribution) lines. During the periods of nominal dust activity, not much differences are observed between the two fitted curves. During the MY 34 GDS and regional dust storm periods (shaded regions), however, the fit that includes the dust shows clear enhancement and agrees better with the observations compared to the one without the dust. The RMS values of the original data, the fit without dust and the fit with dust are 8.45, 7.28, and 8.03, respectively. Thus, the solar flux and solar insolation can account for 86.15% of the variability. With inclusion of the dust, the three together can account for 94.67% of the variability. During the MY 34 GDS, the RMS values of the original data, the fit without dust and the fit with dust are 14.02, 9.94, and 13.17, respectively. Thus, during the GDS the fit without dust accounts for 70.89% of variability in the original data, while the fit with dust accounts for 93.93% of variability.

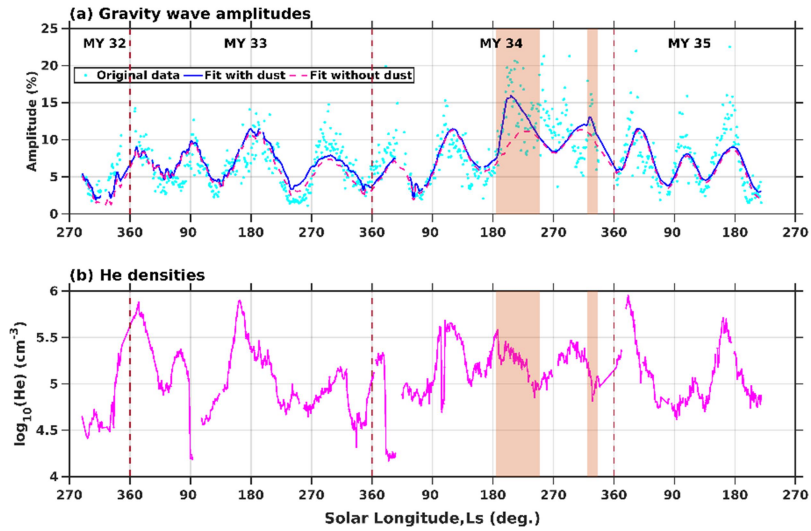

*Figure 5.* Temporal variation of (filled circles) GW amplitudes along with best fit curves (solid) with dust contributions included and (dashed) without the dust contribution. Panel (b) shows the temporal variation of helium (He) densities for an altitude of 200 km. Dashed red vertical lines enclose different Martian years as indicated in the top panel. The shaded regions enclose the approximate periods of the 2018 GDS (Ls=185°-250°) and regional dust storm (Ls=320°-336°).

With inclusion of the dust contribution, equation (4) captures most of the variability in thermospheric GWs. However, we can note that the best fits often overestimate or underestimate the actual measurements. Further insight into these discrepancies can be obtained by comparing them with the helium (He) densities (Figure 5b). Here He is used as a tracer of winds and global circulation since the lighter species are more affected by the winds and circulation than the heavier species (Bougher et al. 2015; Elrod et al. 2017). In general, the discrepancies between the observed and fitted (dust contribution included) GWs are more when the He densities are at their extreme high or low. The discrepancies are particularly large during and immediately after the peak of the MY 34 GDS. This probably indicates that the winds and circulation of the thermosphere and that of the underlying atmosphere partly contribute to the variability of the thermospheric GWs. Note, however, that there are also times when the larger discrepancies between the fitted and original GW amplitudes are not coinciding with the extreme high or low values of He (Figure 5).

The fourth term of equation (4) constrains the seasonal, latitudinal, and diurnal variations to be interdependent. These variabilities are further detailed in Figures 6 & 7. Figure 6 shows the Ls vs. latitude variation of the GW amplitudes corresponding to the fourth term of equation (4) on the nightside (top panel) and dayside (bottom panel). Note that this term adds up to the remaining terms in equation (4). Hence, the positive values of the GW amplitudes add up to the amplitudes accounted for by the other terms in equation (4) and vice versa. On the nightside, the GW amplitudes are enhanced and the largest enhancement is observed in the southern hemisphere at the aphelion. On the dayside, the amplitudes are reduced and the strongest reduction occurs at the perihelion in the southern hemisphere. This variation in GW amplitudes is nearly opposite to that of the solar insolation. At any given latitude and LST, the seasonal variation in GW amplitudes is not more than 4 % and at any given Ls and LST, the latitudinal variation is not more than 6%. Figure 7 shows the LST vs. latitude variation of the solar-insolation related GW amplitudes in the four seasons. In equinoxes, the strongest diurnal variation (8-9%) in GW amplitudes is observed over the equator and the diurnal variation is minimum at the polar regions. In addition, the amplitudes are minimum in the afternoon hours and are maximum in the post-midnight. The strongest GW amplitudes observed at the aphelion (Ls=90°) post-midnight hours are ~10% greater than the weakest GWs observed at the perihelion (Ls=270°) afternoon hours. In general, the diurnal variability in solstices lies between 4% and 6%.

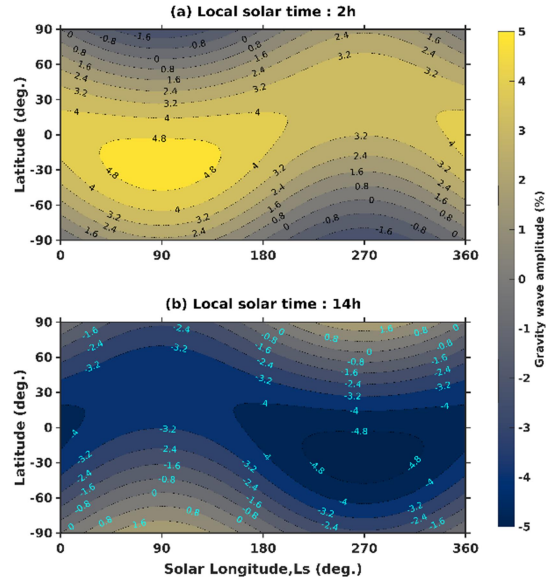

*Figure 6.* Ls vs. latitude variation of the GW amplitudes driven by solar insolation (the fourth term of equation 4) at (a) night time (2 h LST) and (b) daytime (14 h LST).

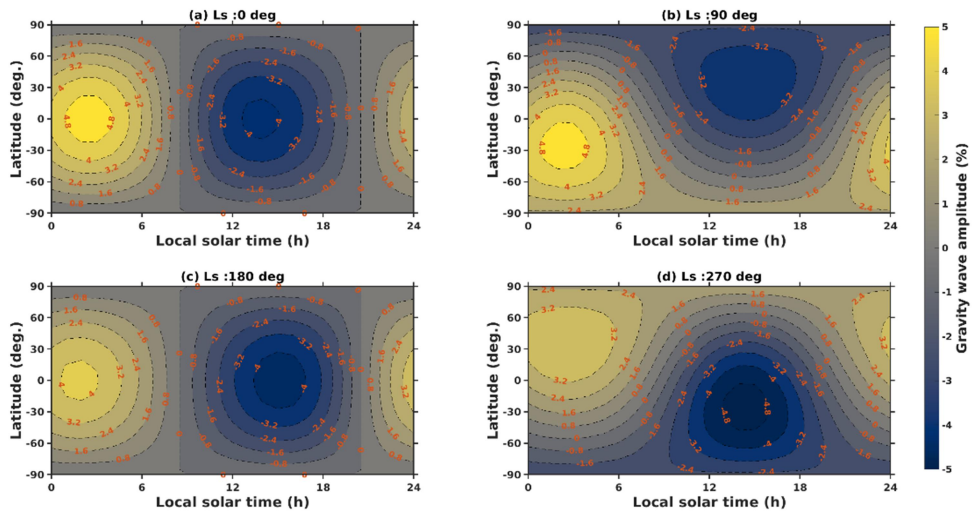

*Figure 7.* Ls vs. latitude variation of the GW amplitudes due to the solar insolation term alone (corresponding to the fourth term of equation 4) for (a-d) Ls=0°, 90°, 180°, and 270°, respectively.

In the following, we use the coefficients of solar activity and the τ of equation (4) to obtain the GW amplitudes driven by these two variables from MY 24, the year from which τ data are available, until MY 35. Figure 8a shows the solar activity (the solar flux corrected to a distance of 1.66 AU) and τ from the beginning of MY 24 to the end of MY 35. The data spans nearly two solar cycles. During this period there were three GDSs; in MY 25 (2001), MY 28 (2007) and MY 34 (2018) and one regional dust storm in MY 34 (Ls=220°-336°). By comparing the solar activity and τ, it can be noted that the MY 28 and MY 34 GDSs occurred during the solar minimum whereas the MY 25 GDS occurred in the solar maximum. Figure 8b and 8c show the GW amplitudes driven by solar activity and τ. The high solar activity during the solar maximum can reduce the GW amplitudes up to ~8% and during the solar minimum the reduction is ~2%. The amplitudes of the dust driven GWs are as high as 5% during MY 25 and MY 34 GDS and are ~3% during the MY 28 GDS. Note that these values are based on the coefficients obtained using the data averaged over one Ls (nearly 8-9 MAVEN orbits). Therefore, instantaneous values of the GW amplitudes driven by these two forcings are expected to be much larger.

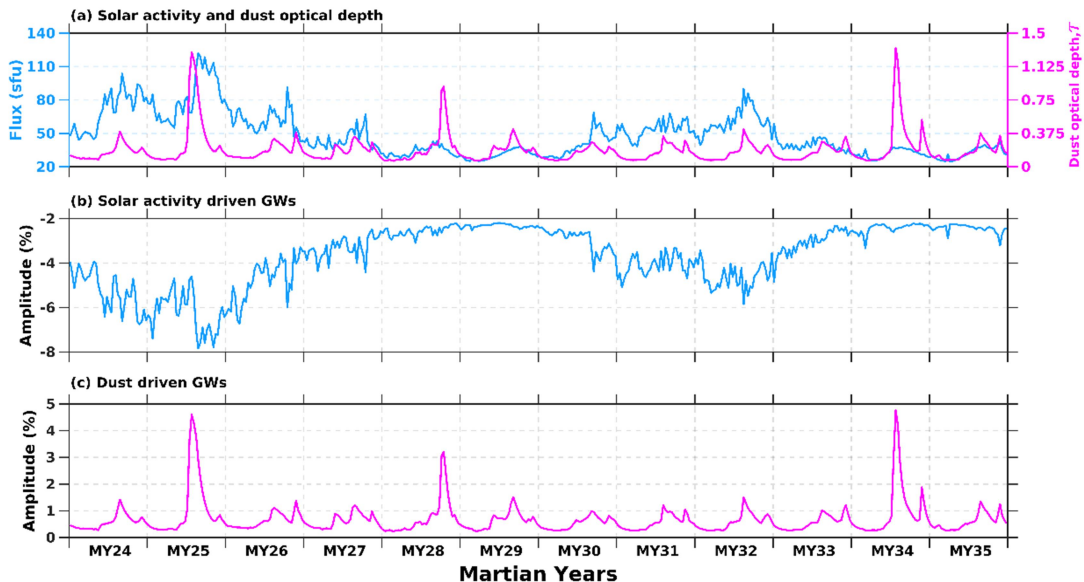

*Figure 8.* Temporal variation of (a) solar activity and the dust optical depth (τ), and the amplitudes of GWs driven by (b) solar activity and (c) the lower atmospheric dust. The dust optical depths are averaged over all longitudes and between ±60° latitudes.

Figures 9 and 10 show the Ls vs. latitude variation of the thermospheric GW amplitudes for nighttime (2 h) and daytime (14 h), respectively for all Martian years from MY 24 to MY 35. Note that these GW amplitudes are calculated using all the four coefficients of equation (4). On the nightside, the GW amplitudes maximize at latitudes that are conjugate to those that receive maximum solar insolation on the dayside. Accordingly, at the aphelion the GW amplitudes are larger in the southern low- and mid-latitudes. At the perihelion, they are larger in the northern low- and mid-latitudes. On the dayside, GW amplitudes are generally smaller, particularly in MY 24 and MY 25, which correspond to the solar maximum of solar cycle 23. Larger GW amplitudes are observed in the winter polar regions. Superposed on this, the GDS related enhancements in the GW amplitudes during MY 25, MY 28, and MY 34 are apparent. The dust related enhancements are mostly confined to ±60° latitudes with a slight shift towards the southern hemisphere in accordance with the latitudinal distribution of the dust in the lower atmosphere (Montabone et al. 2020).

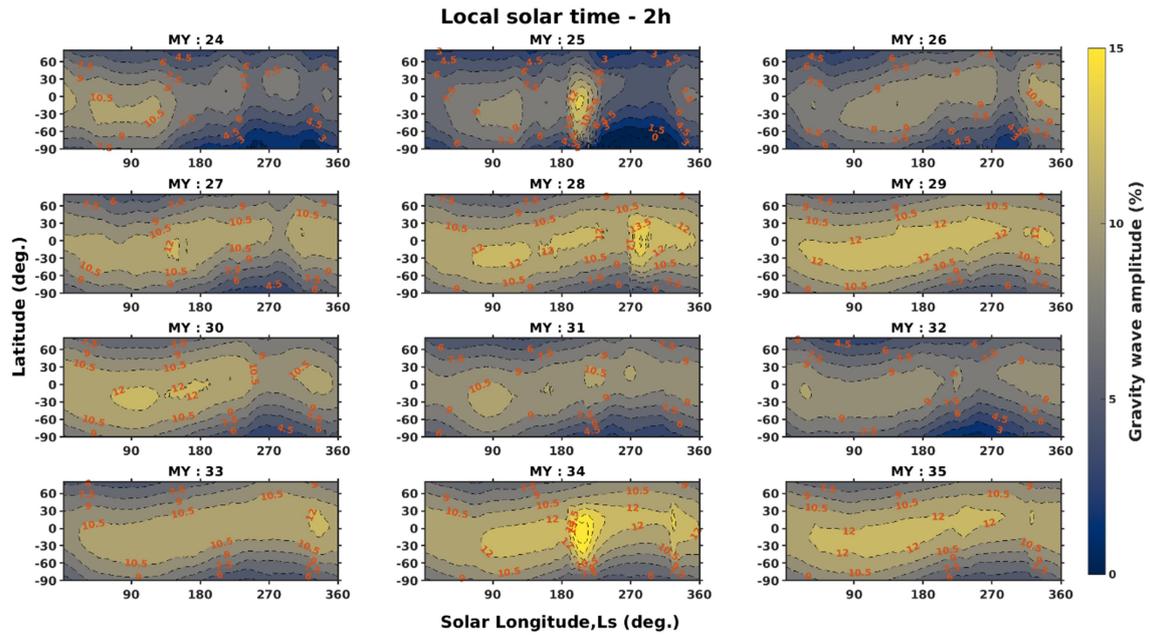

*Figure 9:* *Ls vs. latitude variation of the predicted thermospheric GW amplitudes for nighttime (2 h LST) using all the four regression coefficients of equation (4) and the historical data of solar activity and τ from MY 24 to MY 35. The τ values are averaged over all longitudes and 5° latitude. Note that the solar activity was maximum in MY 24-25 and MY 32 and minimum in MY 28-29 and MY 34-35. Global dust storms occurred in MY 25, MY 28, and MY 34. Ls= Solar longitude.*

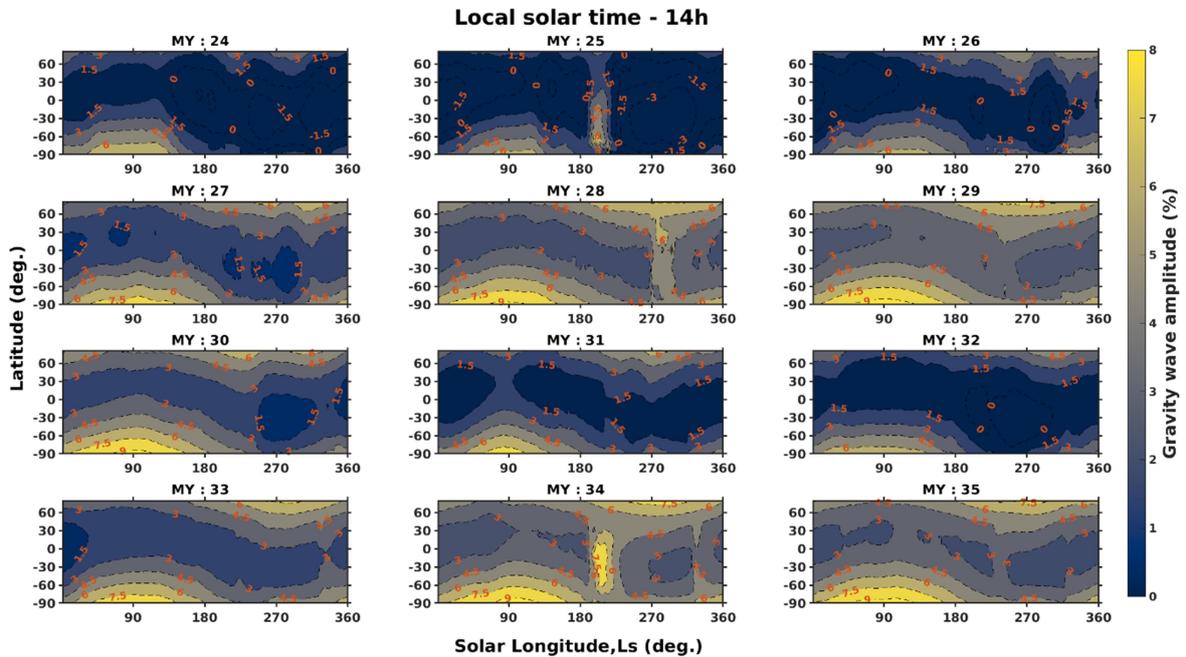

*Figure 10. Same as Figure 9, but for daytime (14 h LST)*

## 4. DISCUSSION

Spacecraft observations for the past couple of decades have been showing compelling evidence that the primary variability in the thermospheric GWs is driven mainly by changes in the underlying atmosphere. Thermal and dynamical state of the underlying atmosphere appears to regulate the amplitudes of the upward propagating GWs. In this study, we extracted the amplitudes of the GWs in the Martian thermosphere from the neutral densities measured by NGIMS/MAVEN. The amplitudes show a complex variability pattern. An examination of these data suggests that the solar activity, solar insolation and the lower atmospheric dust are the three important factors that regulate the amplitudes of the thermospheric GWs locally. A regression analysis has been employed to disentangle the complex variabilities in the GW amplitudes and to isolate the contributions of the three dominant drivers. The residual variabilities are likely driven by winds and circulation of the underlying atmosphere or are associated with the variability at the sources of GW generation. The longest dataset of MAVEN spanning nearly more than five Earth years and the slow precession of the spacecraft's periapsis in latitude and LST have been pivotal in successfully isolating the contributions of the dominant drivers.

The results of the present study show that for a 100 K rise in thermospheric temperature, the GW amplitudes decrease by ~6% (when averaged over a Martian sol). Such an inverse relation has been qualitatively shown in previous studies (England et al. 2017; Terada et al. 2017; Vals et al. 2019; Rao et al. 2021). Stone et al. (2018) have shown that the thermospheric temperatures have a diurnal variability of 80-100 K. This implies that we can expect a maximum of 6% diurnal change in the GW amplitudes. Thermospheric temperatures vary by ~150 K from solar minimum to solar maximum (Bougher et al. 2015; González-Galindo et al. 2015), implying ~9% corresponding change in GW amplitudes.

Using the methodology developed in this study, we show that a 100 sfu increase in the solar activity (the F10.7 cm solar flux corrected for a constant distance of 1.66 AU) can decrease the amplitudes of the thermospheric GWs by ~9%. Note that this decrease is solely due to the increase in the solar activity at the Sun and not due to the change in the Sun-Mars distance. Even at the solar minimum, the solar activity can reduce the GW amplitudes by ~2%. Using NGIMS observations, Rao et al. (2021) and Yiğit et al. (2021b) have come up with nearly similar conclusions that the GW amplitudes are larger during the solar minimum compared to those at solar moderate conditions. Though these authors attempted to quantify the solar flux dependence, the complex interplay of several factors (both in terms of measurements and variability of the drivers) limited such an attempt. The present study addressed this gap.

The results of the present study further show that the spatial distribution and temporal variation of the solar insolation lead to the diurnal, latitudinal, and seasonal variations in GW amplitudes. In fact, the three variations are interdependent and are addressed together through a single term (the fourth term in equation (4)) in the regression analysis. To the knowledge of the authors, this is the first study that appropriately addresses this aspect. Though the seasonal, diurnal and latitudinal variations are coupled, some assessment on the individual variations is made in this study. In general, diurnal variations are stronger (a maximum of 8-9%), followed by latitudinal (~6%) and seasonal (~4%). Larger GW amplitudes during the nighttime and smaller amplitude during daytime reported in previous studies (Yiğit et al. 2015; England et al. 2017; Terada et al. 2017; Vals et al. 2019; Rao et al. 2021) is a direct consequence of the stronger diurnal variation. On the other hand, the weaker amplitudes of the latitudinal and seasonal variations might be the reason why these variabilities were not directly apparent in the previous studies.

The results of the present study show that the dust driven GW amplitudes constitute an important part of the GW variability, particularly during GDSs. The present study shows that for one unit increase in τ, the GW amplitudes increase by 4-6% (when averaged over one Ls). In fact, Leelavathi et al. (2020) and Yiğit et al. (2021a) have also shown such an increase during the 2018 GDS, where the instantaneous amplitudes reached as high as 40% of the background densities. It is interesting to note that among the expected sources of the GWs (both orographic and non-orographic), only the dust storm signatures are clearly apparent in the thermosphere. However, the exact mechanism by which dust enhances the GW amplitudes in the thermosphere is not yet clear. It may be noted that GW amplitudes are damped in the lower atmosphere during dust storms (Heavens et al. 2020; Kuroda et al. 2020). The enhanced GWs in the thermosphere have implications for the transport of water to the upper atmosphere and to escape of hydrogen to the outer space (Yiğit et al. 2021a; Shaposhnikov et al. 2022). Yiğit et al. (2021a) have shown that a 40% increase in GW amplitudes doubles the hydrogen escape flux through the thermal escape mechanism. Shaposhnikov et al. (2022) have shown that the GWs alter the timing and intensity of the water transport to the upper atmosphere. Therefore, the analysis of the present study that disentangles the complex variabilities in the GW amplitudes and isolates the contributions of different drivers will be helpful to assess the role of GWs on the water transport and on the hydrogen escape under different solar, dust and seasonal conditions.

While the solar activity, solar insolation and the lower atmospheric dust could account for most of the GW variability in the thermosphere, it is also important to understand the residuals. The results of the present study show that the fit that is constructed by considering the contributions of the above three drivers deviates more from the measured values when the He densities measured locally are at their extreme high or low. Note that the winds and global circulation of the Martian thermosphere drive the lighter species (such as H, He) more efficiently than the heavier species (such as Ar, $CO_2$) (Elrod et al. 2017; Gupta et al. 2021). As a result, divergence and decrease of He occurs on the dayside and in the local summer and convergence and bulging of He occurs on the nightside and in the local winter. Thus, the extreme high and low values of He are caused by winds and global circulation. The fact that the best fit curve based on equation (4) deviates more when the He densities are at their extreme high or low indicates that the discrepancies in GW amplitudes are likely caused by the effects of winds and circulation, particularly in the mesosphere and lower thermosphere. The discrepancies are particularly large during and immediately after the peak of MY 34 GDS. This further supports the idea that the MY 34 GDS altered the winds and the circulation pattern in the underlying atmosphere that resulted in the background conditions conducive for the upward propagation of the GWs, leading to their enhancement in the thermosphere (Leelavathi et al. 2020). However, it should be noted that from the spacecraft measurements of He, we can only get the basic view, but not the exact picture, of the changes in the winds and circulation pattern (Gupta et al. 2021). This may be the reason why the large discrepancies between the fitted and measured GWs are often not coinciding with the extreme high or low values of He (Figure 5). This suggests that the causes of the discrepancies can be better understood with the knowledge of the exact pattern of the upper atmospheric circulation. Quantification of the circulation effects using regression analysis is not attempted in the present study due to the complexity involved in the expected meridional circulation, which is characterized by a two-cell pattern in equinoxes and one-cell pattern in solstices (Bougher et al. 2015; Elrod et al. 2017). Moreover, the convergence and divergence regions of the winds are themselves variable and often depart from the expected behaviour (Gupta et al. 2021) leading to complex spatial distribution of the adiabatic heating and cooling.

## 5. SUMMARY AND CONCLUSIONS

NGIMS/MAVEN observations of neutral densities are used to obtain the GW amplitudes and temperatures of the Martian thermosphere. The NGIMS measurements constitute the longest data of the Mars thermospheric neutral densities. These data were obtained during the medium to low solar activity period and when the Mars lower atmosphere witnessed the global and regional dust storms that perturbed the thermosphere significantly (Elrod et al. 2019; Jain et al. 2020; Venkateswara Rao et al. 2020). The main results of our study are as follows:

1. The NGIMS observations have shown clear signatures of the effects of the solar activity, solar insolation and the lower atmospheric dust on the thermospheric GWs.
2. A linear regression analysis is used to disentangle the complex variabilities in the thermospheric GWs and to quantify the contributions of the dominant drivers. The solar insolation term that is employed in the regression equation is crucial in this approach.
3. The GW amplitudes show an inverse relation to the thermospheric temperatures and to the solar activity. There is ~6% decrease in GW amplitudes for every 100 K rise in thermospheric temperatures. Similarly, there is ~9% decrease in the GW amplitudes for 100 sfu increase in the solar activity. The solar activity reduces the GW amplitudes by ~2% at solar minimum and ~8% at solar maximum.
4. The solar insolation term constrains the latitudinal, seasonal and diurnal variations to be interdependent. The maximum values of these variations are 6%, 4%, and 9%, respectively.
5. One unit change in the lower atmospheric dust leads to ~4% increase in GW amplitudes (when averaged over one Ls). Accordingly, the GDSs in MY 25 and MY 34 increased the GW amplitudes by ~5% whereas the MY 28 GDS increased them by ~3%.
6. The solar activity, solar insolation, and the lower atmospheric dust put together explain ~85-90% of variability observed in the thermospheric GWs. Remaining 10-15% variability is likely due to the changes in the atmospheric circulation or due to the variability at the GW sources.

The present study, thus, disentangles the complex variabilities in the GW amplitudes in the Martian thermosphere and isolates the contributions of the dominant drivers. Though the regression analysis used in this study could successfully quantify the solar and dust related effects, the role of underlying circulation and the variability at the sources need to be properly addressed. This demands further observations and model simulations. It is important to note that the instantaneous amplitudes of the GWs in the Martian thermosphere are as high as 50% of the background densities which may have implications for spacecraft aerobraking maneuvers. From the results of the present study, it can be inferred that the spacecraft are less likely to experience stronger fluctuations if the maneuvers are carried out on the dayside, during summer time and in the solar maximum period. On the other hand, the spacecraft may experience larger fluctuations if the maneuvers are done on the nightside, from the winter poles and in the solar minimum.

**ACKNOWLEDGEMENTS**

The NGIMS/MAVEN data used in this study are publicly available at the MAVEN Science Data Center (https://lasp.colorado.edu/maven/sdc/public/pages/datasets/ngims.html) and the NASA Planetary Data System (https://atmos.nmsu.edu/data_and_services/atmospheres_data/MAVEN/ngims.html). The maps of gridded and kriged CDOD used in this work have been downloaded from http://www-mars.lmd.jussieu.fr/mars/dust_climatology/index.html. The F10.7 cm radio flux have been taken from https://lasp.colorado.edu/lisird/data/penticton_radio_flux/. V. Leelavathi thanks the Department of Science and Technology, Government of India for supporting her research through INSPIRE fellowship.